\documentclass[10pt,conference]{IEEEtran}
\IEEEoverridecommandlockouts
\usepackage{cite}
\usepackage{amsmath,amssymb,amsfonts}
\usepackage{textcomp}
\usepackage{xcolor}
\usepackage{physics}
\usepackage{graphicx}
\usepackage{tabularx}
\usepackage{dcolumn,booktabs}
\usepackage{subcaption}
\usepackage{threeparttable}
\usepackage[ruled,vlined,linesnumberedhidden]{algorithm2e}
\def\BibTeX{{\rm B\kern-.05em{\sc i\kern-.025em b}\kern-.08em
    T\kern-.1667em\lower.7ex\hbox{E}\kern-.125emX}}

\begin{document}
\bstctlcite{IEEEexample:BSTcontrol}
\title{
Parallelizing quantum simulation with decision diagrams
}

\author{\IEEEauthorblockN{Shaowen Li\IEEEauthorrefmark{1},
Yusuke Kimura\IEEEauthorrefmark{2}, Hiroyuki Sato\IEEEauthorrefmark{1},Junwei Yu\IEEEauthorrefmark{1} and
Masahiro Fujita\IEEEauthorrefmark{3}}
\IEEEauthorblockA{\textit{The University of Tokyo, Japan\IEEEauthorrefmark{1}\IEEEauthorrefmark{3}}}
\IEEEauthorblockA{\textit{Fujitsu Limited, Japan}\IEEEauthorrefmark{2}\\
\{li-shaowen879,schuko,yujw\}@satolab.itc.u-tokyo.ac.jp\IEEEauthorrefmark{1},
yusuke-kimura@fujitsu.com\IEEEauthorrefmark{2},
fujita@ee.t.u-tokyo.ac.jp\IEEEauthorrefmark{3}}}

\maketitle

\begin{abstract}
Recent technological advancements show promise in leveraging quantum mechanical phenomena for computation. This brings substantial speed-ups to problems that are once considered to be intractable in the classical world.  However, the physical realization of quantum computers is still far away from us, and a majority of research work is done using quantum simulators running on classical computers. Classical computers face a critical obstacle in simulating quantum algorithms. Quantum states reside in a Hilbert space whose size grows exponentially to the number of subsystems, i.e., qubits.  As a result, the straightforward statevector approach does not scale due to the exponential growth of the memory requirement. Decision diagrams have gained attention in recent years for representing quantum states and operations in quantum simulations. The main advantage of this approach is its ability to exploit redundancy. However, mainstream quantum simulators still rely on statevectors or tensor networks. We consider the absence of decision diagrams due to the lack of parallelization strategies.  This work explores several strategies for parallelizing decision diagram operations, specifically for quantum simulations. We propose optimal parallelization strategies. Based on the experiment results, our parallelization strategy achieves a 2-3 times faster simulation of Grover's algorithm and random circuits than the state-of-the-art single-thread DD-based simulator DDSIM. 
\end{abstract}


\section{Introduction}
In 2019, Google demonstrated quantum supremacy using its Sycamore processor, which can support 53 qubits\cite{48651}. Since then, quantum computation has attracted more attention and has become a promising solution in various fields for solving previously intractable problems in the classical world\cite{Emani_2021}\cite{https://doi.org/10.48550/arxiv.2210.17460}\cite{https://doi.org/10.48550/arxiv.2012.04473}. However,  since realizing physical quantum devices requires a massive investment of resources,  they are still not accessible to most researchers. Currently, most still rely on classical devices for simulating quantum computations. The statevector based simulation, i.e., using matrices and arrays to store quantum operators and quantum states, works fine for mathematical calculations. However, when implemented on classical machines, this method requires an exponentially large memory size, limiting the number of qubits algorithms can support. 

Quantum algorithms are expressed as a unitary evolution of a quantum state. Quantum states and quantum evolution are respectively modeled using vectors and matrices in a Hilbert space whose sizes grow exponentially with respect to the number of subsystems, i.e., qubits\cite{10.5555/1972505}. A straightforward method to simulate quantum algorithms is to store quantum states and operators as one- and two-dimensional arrays. The exponential growth of the array size limits the simulation scale. Binary decision diagrams(BDDs) are a canonical tool for solving the state explosion in model-checking and formal verification\cite{bddperf}. They have been engineered to simulate quantum circuits in the works such as in\cite{https://doi.org/10.48550/arxiv.2007.09304}. In BDDs, nodes represent matrices or sub-matrices, and edges represent quantum state transitions. It is memory-friendly in the sense that identical sub-matrices across all matrices can be stored using a unique node. Variants of BDDs have been proposed to date for simulation and analysis of quantum computing?for example, QuIDD\cite{1269084} and QDD\cite{1656898}. They extend the original BDDs by assigning weights to edges and attaching more terminals. The discussion in this paper is based on QMDDs\cite{1623982}, which explicitly support complex-valued entries and multiple-valued basis states.

Parallelization has proven its success in most large-scale scientific computations. Undoubtedly, it is also a potent tool in quantum simulation. Parallelizing simulations based on statevectors and tensor networks are discussed in\cite{c4723ee3ceda47f98c037b4677692e4f}\cite{https://doi.org/10.48550/arxiv.quant-ph/9804039}. However, it is yet to be answered how decision diagrams can be efficiently parallelized for quantum simulation. We consider this absence due to its distinctive characteristics. First, decision diagrams are designed to avoid data duplication. This leads to a typical time-space trade-off as synchronizations become essential for updates. Second, parallelization strategies for auxiliary data structures in decision diagram packages, such as the operation cache and the unique table, are not unique and have different performance characteristics in various contexts. Third, parallelism can be achieved via different concurrency primitives\ref{9045711}. Their performance requires examination.

This paper answers these questions by investigating different strategies for parallelizing QMDD-based quantum simulation on a shared-memory machine. We examine three concurrency primitives: \emph{tasks}, \emph{threads}, and \emph{fibers}. Fibers are common in building game engines. However, we have yet to see their appearance in the field of quantum simulation. Our experiment results demonstrate its power in parallelizing decision diagrams. Meanwhile, the effects of the operation cache and the unique table are examined, revealing results not suggested before. For example, global and local caches perform distinctively for different numbers of threads and qubits. We summarize our findings: DD-based quantum simulation can be parallelized to boost performance; however, the parallelization strategies require careful examination. Fibers have advantages over threads for quantum simulations. When the number of qubits is large(e.g., above 30), using fibers with thread local operation caches beats other approaches. Fibers with a global operation cache are preferable for an intermediate number of qubits(e.g., 20 to 30). For less than 10 qubits, parallelization brings extra overheads, which may outweigh its benefits. In the extreme case when the circuit becomes highly random, the operation cache can be removed. Our strategy achieves a 2-3 times faster simulation of Grover's algorithm and random circuits than the state-of-the-art single-thread DD simulator DDSIM\cite{9794651}.

This paper is organized as follows. Section~\ref{sec:design} covers the design of our parallel simulation strategy. Section~\ref{sec:experiment} provides the experiments result and section~\ref{sec:conclusion} concludes this paper.

\begin{figure}[t]
  \centering
\includegraphics[height = 5cm, width=9cm]{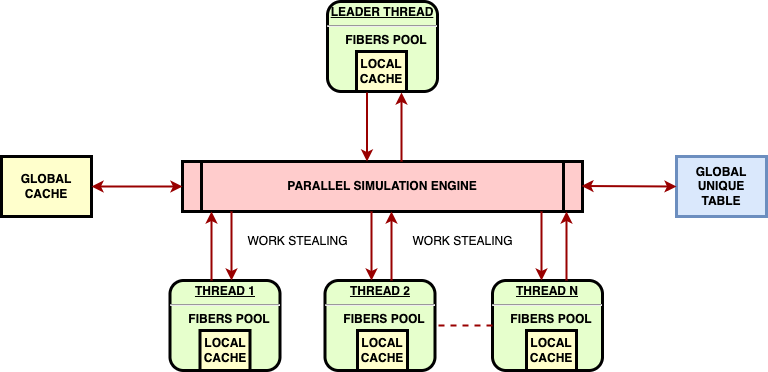}
\caption{Proposed architecture.}
\label{fig:arch}
\end{figure}

\section{Design of Simulation Engine}
\label{sec:design}

The operation cache and unique table are two core components of decision diagram libraries. They are the source of DDs' time and space efficiency, respectively. The unique table ensures that a unique node represents identical sub-matrices.  Thanks to the uniqueness of nodes, calculation results can also be cached. In a parallel simulation engine, the first question to answer is how to synchronize these data structures with the least sacrifice of performance. Furthermore, what parallelization schema most appropriately fits quantum simulation? We propose the architecture in Fig~\ref{fig:arch}. In general, we find fibers have better performance than threads- and tasks-based parallelization. The unique table needs to be global, whereas the choice between a global and local cache depends on the simulation tasks. Work stealing can offer automatic load balancing. Our experiments show this architecture makes better use of resources and accelerates simulations by 2 to 3 times than a single-threaded DD-based simulation. This section illustrates each design choice.

\subsection {Thread local versus global unique table}

\begin{figure}[t]
  \centering
\includegraphics[width=8cm, height = 4cm]{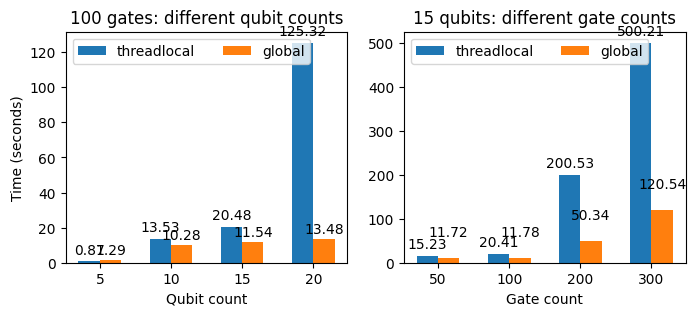}
\caption{A comparison between thread local and global unique table.}
\label{fig:comp}
\end{figure}

Decision diagrams achieve a higher memory efficiency by using the unique table. A single copy of identical nodes is stored in the unique table and shared among decision diagrams. We can choose the unique table to be thread local or global. A thread local unique table can be accessed with no synchronization. However, it reduces the effects of memory saving as the same node not present in the local table may appear in other tables.  The fundamental issue of using a local unique table is that it renders the operation cache hard to hit. The operation cache uses the memory addresses of operand nodes for equality checking, and if the result is present in the cache, the complete computation can be skipped. Using thread local unique tables generates different addresses for the same nodes if they are created by different threads. This does not make the computation inaccurate; however, the cache hit ratio dramatically drops. Fig.~\ref{fig:comp} compares a thread local and global unique table in a random circuit. The circuit consists of 100 gates randomly sampled from the set of \emph{RX, RY, RZ}, and \emph{CNOT}. We conclude that a thread local unique table is inferior to a global unique table for both a higher qubit count and depth. The former suffers from a considerable drop in the cache hit ratio. For example, in a circuit with 100 gates and 20 qubits, the multiplication operation cache hit ratio drops from  $17.54\%$ to nearly none. A global unique table is used in the rest of the comparison. The synchronization is managed by updating the pointer pointing to the next entry in the linked list using \emph{compare-and-swap}.

\subsection {Thread local versus global operation cache}

Similar to the unique table, the operation cache can be either global or local. A global operation cache shares calculation results conducted by any thread with others, whereas a local operation cache only serves its owner. The global cache can theoretically cache more computations with an additional cost of synchronization. However, quantum simulations exhibit both time and spatial locality. First, besides some circuit identities, most quantum gates do not commute and thus must be applied in sequence. Furthermore, each quantum gate is constructed with tensor products between identity gates and the target gate. This makes the resulting matrix contain many sub-identities that can preserve sub-trees from the input state to the output state.  The locality supports the use of thread local caches. 

\begin{figure}[t]
  \centering
\includegraphics[width=8cm, height = 4cm]{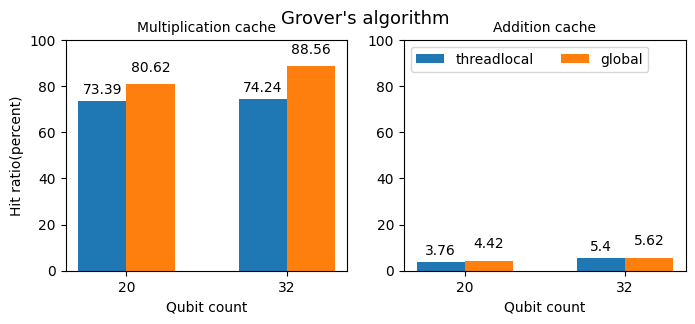}
\caption{A comparison between thread local and global operation cache.}
\label{fig:cache}
\end{figure}

QMDDs divide each matrix into four sub-matrices, and most cache hits happen among the calculations on these sub-matrices since quantum matrices tend to be structured(e.g., symmetric). The majority of these calculations are done within a single thread. This also reinforces the use of thread local caches. We observe that the hit ratio of using thread local caches is only lower than that of a global cache to a slight extent but comes with no cost of synchronization. In fig.~\ref{fig:cache}, we present a comparison in Grover's algorithm. The gap between the local and global cache hit ratio is small. The local cache relieves the burden of synchronization and leads to better performance. We also see that the hit ratio of the multiplication is tens of times higher than the addition cache. The hit ratios for multiplication and addition are virtually zero for a random circuit, so they are not shown. From the cost-effectiveness perspective, assigning a larger memory space to the multiplication cache is more sensible. 

\subsection{Fibers versus threads}

Fibers are a lightweight execution context in the userspace and share some data with the underlying threads. For example, fibers own stacks but share the address space with the underlying thread. This means the same amount of memory can support more fibers, and the context switch among fibers is faster. Fibers use cooperative scheduling. The idea is that they deliberately yield to others at a chosen point. It should be noted that using fibers does not provide extra concurrency. The concurrency level still depends on the number of operating system threads(hardware threads, to be precise). Within a single thread, only one fiber can execute at a single point in time. The benefit of this is that no extra synchronization needs to be handled at the level of fibers as long as threads are appropriately synchronized.

\subsection{Parallelization schema}
We analyze three strategies for parallelizing the simulation process: \emph{task-based outer parallelization}, \emph{thread-based inner parallelization}, and \emph{fiber-based inner parallelization}.

\begin{figure}[t]
\centering

\begin{minipage}[b]{0.45\hsize}
\includegraphics[width=\textwidth]{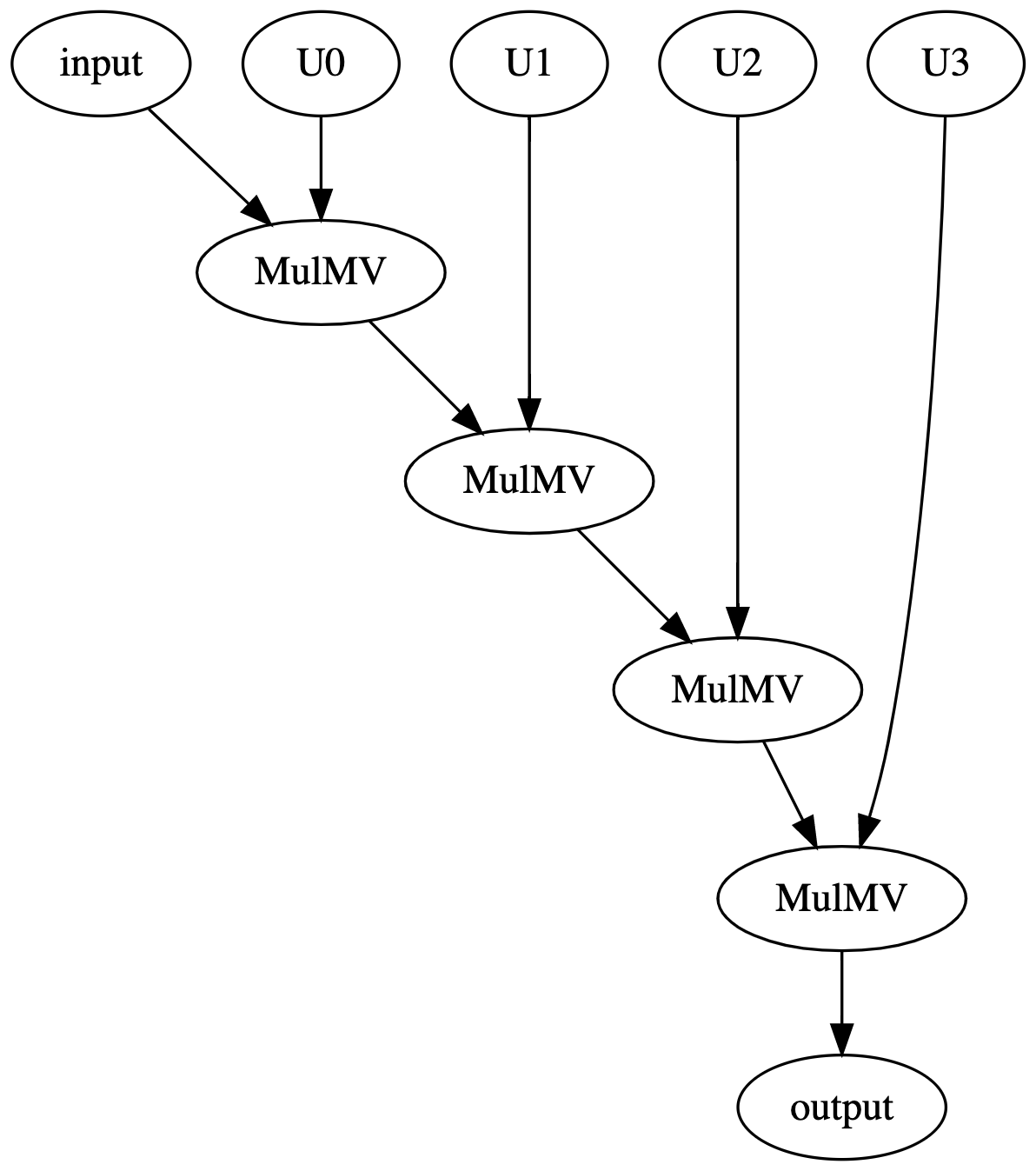}
\subcaption{A linear task graph.}
\label{fig:outer1}
\end{minipage}

     
\begin{minipage}[b]{0.45\hsize}
\includegraphics[width=\textwidth]{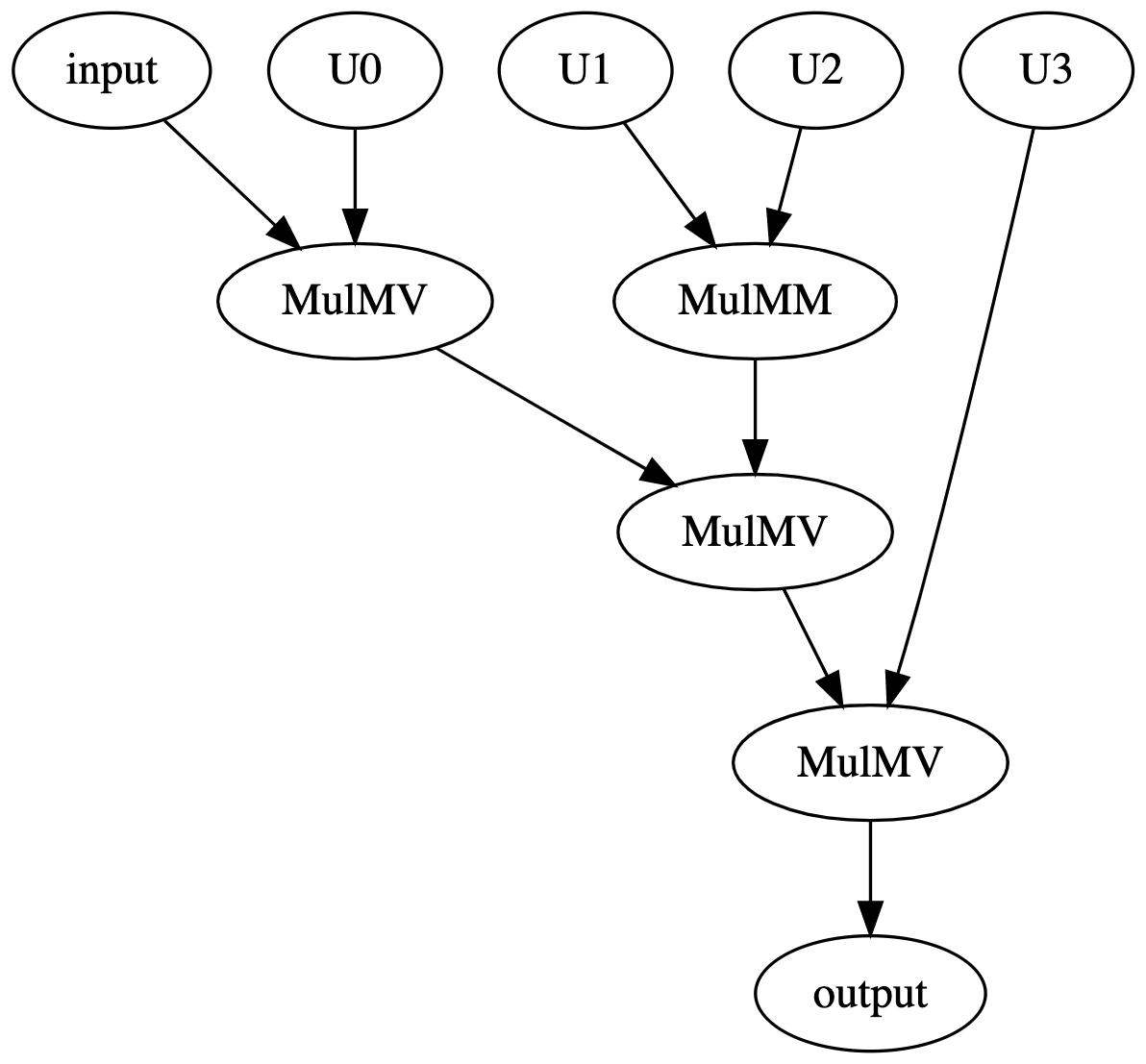}
\subcaption{A associative task graph.}
\label{fig:outer2}
\end{minipage}

\end{figure}

Consider a quantum circuit representing the state evolution of $\ket{\text{output}}=U_4U_3U_2U_1\ket{\text{input}}$. Task-based outer parallelization launches a fixed number of worker threads and creates tasks for each multiplication $U_{i+1}U_i \text{ or } U_i\ket{\text{state}}$. The non-commutativity of multiplication induces dependencies among tasks. Task dependencies are handled by constructing a task graph with nodes representing operations and edges representing task prerequisites. Fig.~\ref{fig:outer1} shows a toy example.In the task graph, we use \emph{MulMV} and \emph{MulMM} to represent the operation of matrix-vector multiplication and matrix-matrix multiplication. Such a task graph does not offer any true concurrency. For example, all \emph{MulMV} nodes must be executed in a linear sequence. We improve this by leveraging the fact that multiplication is associative. Thus worker threads can simulate different parts of the circuit simultaneously. We use associativity to construct the task graph in fig.~\ref{fig:outer2}. Dependencies among \emph{MulMV} nodes are removed, and thus they can be executed in parallel. We call this \emph{outer parallelization} since sub-trees of a single decision diagram are all processed by the same thread. The whole task graph is processed in parallel by multiple threads. Nevertheless, we find this approach performs poorly in our experiment. Matrix-vector multiplication is faster than matrix-matrix multiplication in general. This also applies to decision diagrams. Therefore, it is more efficient to reduce the dimension faster and earlier by allocating more resources(i.e., worker threads) to compute matrix-vector multiplication rather than matrix-matrix multiplication. However, \emph{MulMV} nodes in fig~\ref{fig:outer2}, the only type of node generating a vector and reducing the dimension, it only taken by a single thread.

\begin{figure*}[t]
\centering
\includegraphics[width = 0.5\textwidth]{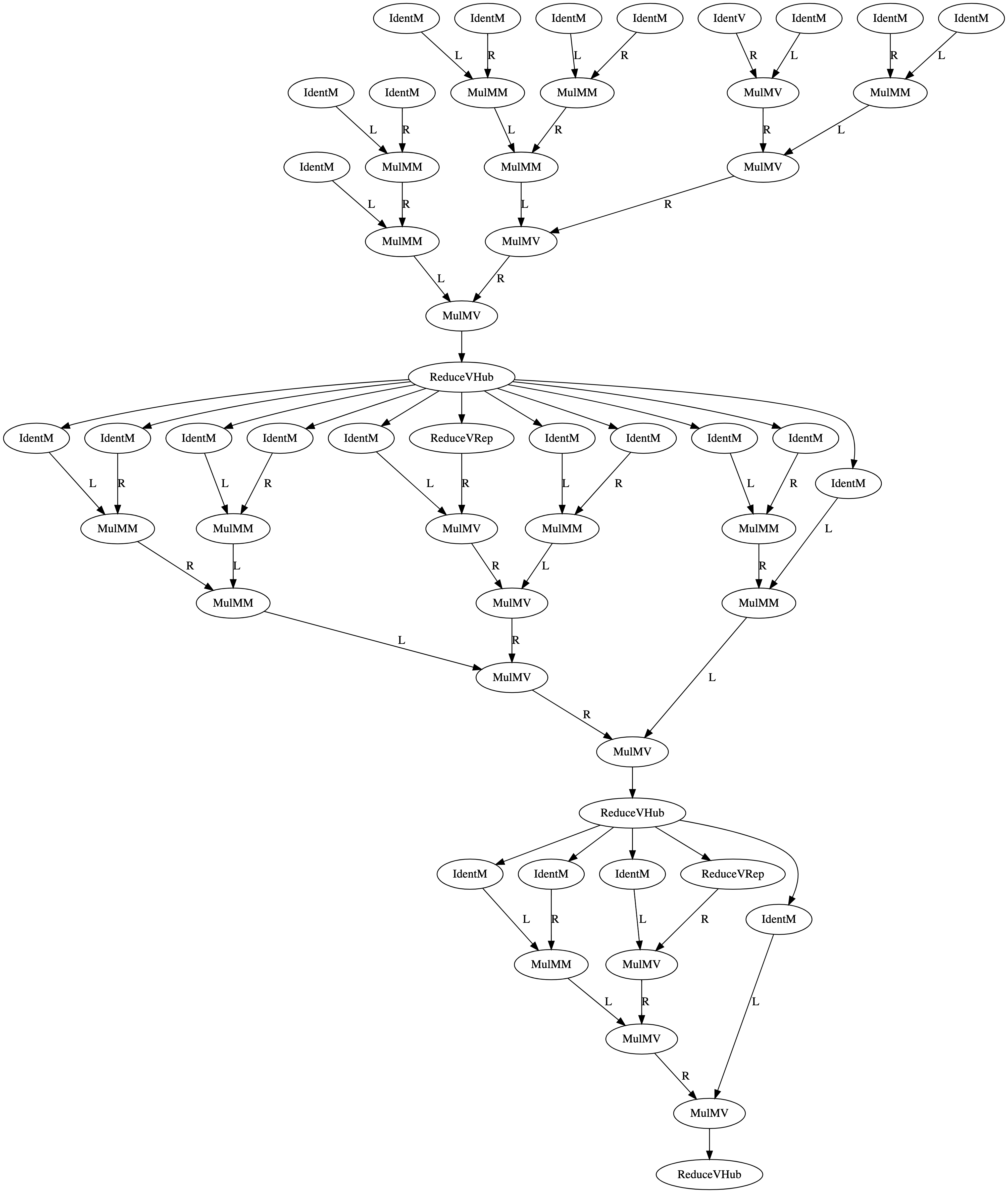}
\caption{A task graph with \emph{reduce} nodes.}
\label{fig:reduce}
\end{figure*}

We recommend constructing a task graph in which dependencies are set such that tasks are processed in batches following the flow of the circuit. This means worker threads work jointly in a sub-part of the circuit to accelerate the dimension reduction and avoid working on arbitrary parts of the circuit. We construct such a task graph by introducing a node of type \emph{Reduce}. It serves as a \emph{hub} in the graph. The \emph{reduce} nodes segregate the graph and enforce worker threads to work on earlier parts of the circuit before later can be processed. Fig.~\ref{fig:reduce} shows an example of such a graph. In our experiment, we find the inclusion of \emph{reduce} nodes  improves the performance by approximately two times compared with the other two kinds of task graph.

\begin{figure}[t]
\centering
\includegraphics[width=8cm, height = 6cm]{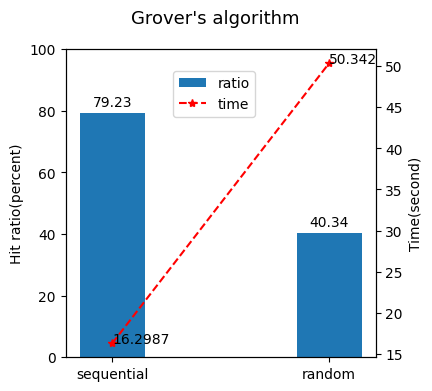}
\caption{Grover's algorithm with different processing orders.}
\label{fig:process}
\end{figure}

The vital problem of the \emph{outer parallelization} is its poor cache utilization. As threads may take arbitrary task nodes, decision diagrams processed by a single thread can come from remote parts of the circuit and thus exhibit low similarities. This means cached results are rarely reused, leading to a lower cache hit ratio. In fig.~\ref{fig:process}, we run Grover's algorithm with a single thread using the same task graph but different processing orders: one for sequentially multiplying decision diagrams and one for picking decision diagrams randomly. It illustrates that random processing decreases the cache hit ratio and leads to a longer execution time.

\begin{table*}[t]
\centering
\begin{threeparttable}[c]
\caption{RESULTS ON  GROVER'S ALGORITHM}
\begin{tabular}{|c|lllr|llr|r|r|}
\hline
\textbf{Qubit} & \multicolumn{4}{c|}{\textbf{Local Cache}}                                                    & \multicolumn{3}{l|}{\textbf{Global Cache}}                            &  &   \\ \hline
          & \multicolumn{1}{c|}{single} & \multicolumn{1}{c|}{fibers} & \multicolumn{1}{c|}{OpenMP} & taskgraph & \multicolumn{1}{c|}{fibers} & \multicolumn{1}{c|}{OpenMP} & taskgraph & DDSIM  & Qiskit Aer  \\ \hline
         10 & \multicolumn{1}{r|}{0.701} & \multicolumn{1}{r|}{1.231} & \multicolumn{1}{r|}{2.451} & 0.122  & \multicolumn{1}{r|}{0.054} & \multicolumn{1}{r|}{0.512} & 0.964 & 0.353 & \textbf{0.053} \\
         12 & \multicolumn{1}{r|}{0.854} & \multicolumn{1}{r|}{3.789} & \multicolumn{1}{r|}{3.732} & 0.255 & \multicolumn{1}{r|}{\textbf{0.044}} & \multicolumn{1}{r|}{0.863} & 0.863 & 0.324 & 1.788 \\
         15 & \multicolumn{1}{r|}{0.833} & \multicolumn{1}{r|}{3.995} & \multicolumn{1}{r|}{4.125} & 1.786 & \multicolumn{1}{r|}{\textbf{0.037}} & \multicolumn{1}{r|}{0.076} & 0.984 & 0.475  & 538.869 \\
         18 & \multicolumn{1}{r|}{0.856} & \multicolumn{1}{r|}{3.865} & \multicolumn{1}{r|}{5.915} & 178.315 & \multicolumn{1}{r|}{\textbf{0.017}} & \multicolumn{1}{r|}{0.187} & 76.253 & 0.466 & OOM \\
         21 & \multicolumn{1}{r|}{1.061} & \multicolumn{1}{r|}{4.124} & \multicolumn{1}{r|}{6.663} & 1756.357& \multicolumn{1}{r|}{\textbf{0.148}} & \multicolumn{1}{r|}{0.474} & 679.356 & 0.874 &  \\
         24 & \multicolumn{1}{r|}{1.223} & \multicolumn{1}{r|}{5.062} & \multicolumn{1}{r|}{7.027} & 5928.232 & \multicolumn{1}{r|}{\textbf{0.368}} & \multicolumn{1}{r|}{1.668} & 4486.656 & 3.129 &  \\
         27 & \multicolumn{1}{r|}{7.654} & \multicolumn{1}{r|}{2.132} & \multicolumn{1}{r|}{8.095} & TIMEOUT\tnote{2} & \multicolumn{1}{r|}{\textbf{1.442}} & \multicolumn{1}{r|}{6.055} & TIMEOUT & 13.254 &  \\
         28 & \multicolumn{1}{r|}{8.734} & \multicolumn{1}{r|}{\textbf{2.967}} & \multicolumn{1}{r|}{10.323} &  & \multicolumn{1}{r|}{3.065} & \multicolumn{1}{r|}{9.346} &  & 19.074 &  \\
         29 & \multicolumn{1}{r|}{10.825} & \multicolumn{1}{r|}{5.075} & \multicolumn{1}{r|}{16.627} &  & \multicolumn{1}{r|}{\textbf{4.869}} & \multicolumn{1}{r|}{21.364} &  & 29.465 &  \\
         30 & \multicolumn{1}{r|}{16.298} & \multicolumn{1}{r|}{\textbf{10.157}} & \multicolumn{1}{r|}{15.486} &  & \multicolumn{1}{r|}{10.422} & \multicolumn{1}{r|}{50.331} &  & 49.016 &  \\
         31 & \multicolumn{1}{r|}{51.677} & \multicolumn{1}{r|}{20.722} & \multicolumn{1}{r|}{107.316} &  & \multicolumn{1}{r|}{\textbf{20.341}} & \multicolumn{1}{r|}{176.749} &  & 62.346 &  \\
         32 & \multicolumn{1}{r|}{94.687} & \multicolumn{1}{r|}{\textbf{50.865}} & \multicolumn{1}{r|}{193.427} &  & \multicolumn{1}{r|}{91.261} & \multicolumn{1}{r|}{385.245} &  & 89.123 &  \\
         33 & \multicolumn{1}{r|}{130.568} & \multicolumn{1}{r|}{\textbf{45.338}} & \multicolumn{1}{r|}{504.925} &  & \multicolumn{1}{r|}{720.365} & \multicolumn{1}{c|}{OOM\tnote{1}} &  & 137.456 &  \\
         34 & \multicolumn{1}{r|}{330.462} & \multicolumn{1}{r|}{\textbf{82.455}} & \multicolumn{1}{c|}{OOM} &  & \multicolumn{1}{r|}{2400.412} & \multicolumn{1}{c|}{OOM} &  & 197.957 &  \\ \hline
\end{tabular}
\begin{tablenotes}
\item [1] Out Of Memory
\item [2] Exceed 7200 seconds 
\end{tablenotes}
\end{threeparttable}
\label{tab1}
\end{table*}

\begin{table*}[t]
\caption{RESULTS ON  Random circuit(depth = 200)}
\begin{center}
\begin{tabular}{|c|lllr|lllr|llr|}
\hline
\textbf{Qubit} & \multicolumn{4}{c|}{\textbf{No Cache}}& \multicolumn{4}{c|}{\textbf{Local Cache}} & \multicolumn{3}{l|}{\textbf{Global Cache}}                              \\ \hline
          & \multicolumn{1}{c|}{single} & \multicolumn{1}{c|}{fibers} & \multicolumn{1}{c|}{OpenMP} & taskgraph &\multicolumn{1}{c|}{single} & \multicolumn{1}{c|}{fibers} & \multicolumn{1}{c|}{OpenMP} & taskgraph & \multicolumn{1}{c|}{fibers} & \multicolumn{1}{c|}{OpenMP} & taskgraph   \\ \hline
         10 &\multicolumn{1}{r|}{0.037} & \multicolumn{1}{r|}{0.023} & \multicolumn{1}{r|}{0.261} & 1.456 & \multicolumn{1}{r|}{0.903} & \multicolumn{1}{r|}{10.551} & \multicolumn{1}{r|}{5.289} & 6.536 & \multicolumn{1}{r|}{0.125} & \multicolumn{1}{r|}{1.851} &12.535  \\
         12 &\multicolumn{1}{r|}{0.152} & \multicolumn{1}{r|}{0.058} & \multicolumn{1}{r|}{1.048} & 2.633 & \multicolumn{1}{r|}{1.059} & \multicolumn{1}{r|}{11.727} & \multicolumn{1}{r|}{6.048} & 10.341 & \multicolumn{1}{r|}{0.437} & \multicolumn{1}{r|}{7.681} & 254.241  \\
         14 &\multicolumn{1}{r|}{0.639} & \multicolumn{1}{r|}{0.159} & \multicolumn{1}{r|}{3.991} & 15.674 & \multicolumn{1}{r|}{1.754} & \multicolumn{1}{r|}{10.965} & \multicolumn{1}{r|}{8.921} & 88.234 & \multicolumn{1}{r|}{1.411} & \multicolumn{1}{r|}{28.847} & 873.436  \\
         16 &\multicolumn{1}{r|}{3.141} & \multicolumn{1}{r|}{\textbf{0.573}} & \multicolumn{1}{r|}{14.733} & 164.634 & \multicolumn{1}{r|}{5.012} & \multicolumn{1}{r|}{12.579} & \multicolumn{1}{r|}{18.786} & 364.523 & \multicolumn{1}{r|}{4.479} & \multicolumn{1}{r|}{95.337} & 2034.231   \\
         18 &\multicolumn{1}{r|}{35.847} & \multicolumn{1}{r|}{\textbf{5.377}} & \multicolumn{1}{r|}{82.552} & 1042.534 & \multicolumn{1}{r|}{42.215} & \multicolumn{1}{r|}{17.649} & \multicolumn{1}{r|}{92.232} & 3442.464 & \multicolumn{1}{r|}{26.091} & \multicolumn{1}{r|}{460.021} & TIMEOUT  \\
         20 &\multicolumn{1}{r|}{147.693} & \multicolumn{1}{r|}{\textbf{21.181}} & \multicolumn{1}{r|}{314.501} & 5654.523 & \multicolumn{1}{r|}{159.452} & \multicolumn{1}{r|}{35.651} & \multicolumn{1}{r|}{305.234} & TIMEOUT & \multicolumn{1}{r|}{78.194} & \multicolumn{1}{r|}{1353.537} &   \\
         22 &\multicolumn{1}{r|}{4804.881} & \multicolumn{1}{r|}{\textbf{631.768}} & \multicolumn{1}{r|}{6226.24} & TIMEOUT & \multicolumn{1}{r|}{4935.112} & \multicolumn{1}{r|}{656.172} & \multicolumn{1}{r|}{6403.212} &  & \multicolumn{1}{r|}{731.654} & \multicolumn{1}{r|}{TIMEOUT} &   \\
         24 &\multicolumn{1}{r|}{TIMEOUT} & \multicolumn{1}{r|}{\textbf{1258.325}} & \multicolumn{1}{r|}{TIMEOUT} &  & \multicolumn{1}{r|}{TIMEOUT} & \multicolumn{1}{r|}{1358.234} & \multicolumn{1}{r|}{TIMEOUT} &  & \multicolumn{1}{r|}{1658.241} & \multicolumn{1}{r|}{} &   \\
         26 &\multicolumn{1}{r|}{} & \multicolumn{1}{r|}{\textbf{1484.643}} & \multicolumn{1}{r|}{} &  & \multicolumn{1}{r|}{} & \multicolumn{1}{r|}{1648.353} & \multicolumn{1}{r|}{} &  & \multicolumn{1}{r|}{1994.234} & \multicolumn{1}{r|}{} &  \\
         28 &\multicolumn{1}{r|}{} & \multicolumn{1}{r|}{\textbf{2346.351}} & \multicolumn{1}{r|}{} &  & \multicolumn{1}{r|}{} & \multicolumn{1}{r|}{2476.423} & \multicolumn{1}{r|}{} &  & \multicolumn{1}{r|}{2857.341} & \multicolumn{1}{r|}{} &   \\
         30 &\multicolumn{1}{r|}{} & \multicolumn{1}{r|}{\textbf{3994.442}} & \multicolumn{1}{r|}{} &  & \multicolumn{1}{r|}{} & \multicolumn{1}{r|}{4124.527} & \multicolumn{1}{r|}{} &  & \multicolumn{1}{r|}{4572.421} & \multicolumn{1}{r|}{} &   \\ \hline
\end{tabular}
\label{tab2}
\end{center}
\end{table*}

\begin{table}[t]
\caption{RESULTS ON  Random circuit continued(depth = 200)}
\begin{center}
\begin{tabular}{|c|r|r|}
\hline
\textbf{Qubit} & DDSIM  & Qiskit Aer  \\ \hline
10 &  0.046  & \textbf{0.022} \\
12 &  0.191  & \textbf{0.025} \\
14 &  6.602  & \textbf{0.037} \\
16 &   55.133 & 1.416 \\
18 &   258.163 & 17.723 \\
20 &   436.237 &  82.552\\
22 &  6942.452  & 725.074 \\
24 &   TIMEOUT & 2593.534\\
26 &    &  OOM\\
28 &    &  \\
30 &    &  \\ \hline
\end{tabular}
\label{tab3}
\end{center}
\end{table}

\emph{Inner parallelization} does not simulate the entire circuit concurrently. Quantum gates are applied to the start state vector in sequence, and only one \emph{main} thread carries out this. What distinguishes this strategy is that different threads will process sub-trees of a single decision diagram. We create fibers when the operation is recursively applied on child nodes until a lower bond of qubit count is reached to avoid having too many fibers.  \emph{Inner parallelization with fibers} offers the following advantages. First, fibers are dynamic, whereas task graphs are static. This means we can query the operation cache first and avoid launching a new fiber if it returns a hit. Second, inner parallelization boosts cache efficiency. Quantum circuits tend to evolve the input state gradually. At each step, quantum gates are applied to the state vector locally. This means a majority of subtrees in the decision diagram remain unchanged across each gate. Third, workloads on different sub-trees are imbalanced due to different tree sizes. Fibers can be inexpensively created, destroyed, and migrated among threads. This realizes an automatic load balancing. In our experiment, we discover that worker threads spend over $50\%$ of time idling and waiting for tasks in \emph{outer parallelization}. In \emph{fiber-base inner parallelization}, this falls to approximately $10\%$.

\section{Experiments}
\label{sec:experiment}
To evaluate the effectiveness of each parallelization strategy, we implement a QMDD-based quantum simulator in C++. We implement our task graph engine. We use \emph{Boost fibers} and \emph{OpenMP}\cite{dagum1998openmp} for \emph{fiber-based} and \emph{thread-based} parallelization. Work-stealing is enabled for load balancing. In the experiments, we compare with the state-of-the-art QMDD-based simulator DDSIM and \emph{Qiskit Aer}\cite{Qiskit} with \emph{default qubit} statevector-based backend. We conduct the experiments on a server with AMD Ryzen 9 7950X 16-core processor and 128GB RAM. The Linux kernel is 5.19.0. We set the timeout limit to 7200 seconds. The Linux kernel manages out-of-memory abortions.

Table~\ref{tab1} presents the result of Grover's algorithm. Our implementation first create the unitary gate for one complete Grover iteration, the number of Grover's iterations is computed by $\lfloor \frac{\pi\sqrt{N}}{4} \rfloor$. The oracle we use can be described by $U\ket{x} = (-1)^{f(x)}\ket{x}$ where $f(x) = 1$ only for a single input. The number of threads is fixed to 16 for all multi-thread cases. Each thread is pinned to a separate core to remove the impacts from thread migration. The results show that \emph{taskgraph} performs poorly for both local and global cache. Substantial overheads come from traversing the task graph and waiting for dependencies. The cache hit ratio of Grover's algorithm is expected to be high because of its repeated applications of Grover's iteration. However, we find the hit ratio is below $10\%$ in \emph{taskgraph} due to worker threads randomly working on different parts of the circuit. This pollutes the cache content and leads to unnecessary evictions. \emph{Qiskit Aer} uses the statevector. It outflows the memory above 18 qubits on our server. It is also slower than all the other decision diagram based approaches. 

When \emph{OpenMP} is used for parallelization, using a global cache accelerates the simulation for small qubit counts. The global cache simulates more than ten times faster than the local cache when the number of qubits is below 24. Local caches start to boost the performance when the qubit count gets larger. We also observe the same phenomenon in the case of fibers. The profiling results suggest the following reasons. First, higher qubit counts generally require a longer execution time. This makes the costs associated with locking the cache more severe.  Second, the size of decision diagrams also increases with qubit count. Therefore, more sub-trees share the same structure, thus the same cache bucket. Using separate local caches alleviates cache contention.

The performance of DDSIM is among the first tier. Our single-thread implementation targets on matching DDSIM to serve as a reasonable baseline. Therefore, their results are comparable. DDSIM uses a single thread. However, it is superior to the sixteen-thread \emph{OpenMP} implementation with a global cache. This confirms the negative impacts of unduly synchronization. In our experiments, when the qubit count is below 27, fibers should be used in conjunction with a global cache. This offers the optimal performance among all strategies we have benchmarked. With more qubits, fibers should be used with local caches. This combination is the only one superseding DDSIM in our experiments, and it reduces the simulation time by 3-4 times.

The issue of a single-thread decision diagram simulator becomes more evident in a random circuit. In table~\ref{tab2} and ~\ref{tab3}, a pure random circuit with depth 200  is tested. We randomly sample 200 gates from the gate set of \emph{RX}, \emph{RY}, \emph{RZ}, and \emph{CNOT}. Each gate's rotation angle and target qubit are also uniformly distributed. Decision diagrams are generally unsuitable for highly random circuits as the operation cache barely helps, and random memory accesses incurred from traversing the decision tree further hurt the performance. Consequently, the \emph{no cache} case performs the best for larger qubit counts. \emph{Global cache} performs worse than \emph{local cache} as it cannot obtain a higher cache hit ratio but adds synchronization overheads.  \emph{Qiskit}'s performance is satisfactory. It is faster than DDSIM for all qubit counts before it outflows the memory. We cannot observe any advantage from using \emph{OpenMP} than a single-thread implementation, suggesting that naive parallelization can, in fact, hurt the performance.

\section{Conclusions and future work}
\label{sec:conclusion}

In this paper, we examined decision diagram based quantum simulation. Using QMDDs as an alternative to the basic binary decision diagrams in quantum simulators has been around for a while. Although its performance is superior to statevectors and tensor networks, its lack of parallelization strategies limits its scale. Components of decision diagram libraries, such as the unique table and operation cache, complicate the adaptation to parallelization. Consequently, traditional thread based strategies bring little, if any, performance improvement. Their synchronization costs occasionally are even harmful to the results. In order to address these problems, we present a comprehensive overview of the trade-offs of several parallelization strategies. Furthermore, we propose a design that allows a faster simulation than the state-of-the-art single-threaded simulator DDSIM.

\bibliographystyle{IEEEtran}
\bibliography{ref}
 
\end{document}